\documentclass[a4paper,fleqn,usenatbib]{mnras}

\usepackage{graphicx}
\usepackage{times}

\usepackage[T1]{fontenc}
\usepackage{ae,aecompl}

\title[Eccentricity distribution of wide binaries]{Eccentricity distribution of wide binaries}

\author[Tokovinin \& Kiyaeva]{A.~Tokovinin$^{1}$\thanks{E-mail: atokovinin@ctio.noao.edu}
\& O.~Kiyaeva$^{2}$\thanks{E-mail: kiyaeva@list.ru} \\
$^1$Cerro Tololo Inter-American Observatory, Casilla 603, La Serena, Chile\\
$^2$Central Astronomical Observatory at Pulkovo, Pulkovskoe sh. 65/1, St. Petersburg 196140, Russia 
}

\date{Accepted XXX. Received YYY; in original form ZZZ}
\pubyear{2015}

\begin{document}
\label{firstpage}
\pagerange{\pageref{firstpage}--\pageref{lastpage}} 

\maketitle

\begin{abstract}
A  sample of  477  solar-type binaries  within  67\,pc with  projected
separations  larger  than  50\,AU  is  studied by  a  new  statistical
method. Speed and direction of the relative motion are determined from
the short observed arcs or  known orbits, and their joint distribution
is compared  to the numerical simulations.  By  inverting the observed
distribution  with  the help  of  simulations,  we  find that  average
eccentricity of  wide binaries  is 0.59$\pm$0.02 and  the eccentricity
distribution can be  modelled as $f(e) \approx 1.2  e + 0.4$.  However,
wide binaries containing inner subsystems, i.e. triple or higher-order
multiples, have significantly  smaller eccentricities with the average
$e=  0.52\pm0.05$ and  the peak  at $e  \sim 0.5$.   We find  that the
catalog   of  visual   orbits   is  strongly   biased  against   large
eccentricities.  A marginal evidence  of eccentricity  increasing with
separation  (or period)  is found  for this  sample.   Comparison with
spectroscopic  binaries  proves   the  reality  of  the  controversial
period-eccentricity relation. The average eccentricity does increase with
binary period, being 0.39 for periods from $10^2$ to $10^3$ days
and 0.59 for the binaries studied here  ($10^5$ to $10^6$ days).   
\end{abstract}


\section{Introduction}
\label{sec:intro}

A  significant fraction  of  stars  are born  in  binary and  multiple
systems.   Statistics  of  their  orbital parameters  bear  traces  of
formation  history and  help understanding  its  physics.  Binary-star
formation is an  essential piece in such fundamental  areas as stellar
mass  function,  formation of  planetary  systems, and  binary/stellar
population synthesis.

Recent  review  of  binary   statistics  by  \citet{DK13}  focuses  on
multiplicity  fractions,  periods,  and  mass ratios,  mentioning  the
eccentricity  distribution  only  briefly.   Yet, eccentricity  is  an
important diagnostic of binary  formation.  It is generally recognised
that dynamical  interactions in stellar systems  produce binaries with
``thermal''  eccentricity  distribution $f(e)  =  2e$, predicted  from
general  considerations by \citet{Ambartsumian}.   On the  other hand,
binaries with mean  eccentricities below 0.5 are  formed in hydro-dynamical
simulations as  a result  of dissipative interaction  with surrounding
gas  which also decreases  binary separations  \citep{Bate2009}.  From
this   perspective,   spectroscopic   binaries   should   have   lower
eccentricities  than wider  visual binaries,  and a  general  trend of
increasing eccentricity with binary period is expected.

The relation between eccentricity and binary period, noted a long time
ago, has been controversial ever since. While tidal circularisation at
periods shorter than $\sim$10  days is well established, the situation
at longer periods is not clear. \citet{DM91} studied a small sample of
nearby solar-type stars and found  that at periods between 10 and 1000
days the  mean eccentricity is 0.31,  same as in  nearby open clusters
and halo. At longer periods, they found a ``bell-shaped'' eccentricity
distribution and  suggested that after correction  of selection biases
the data are compatible with the thermal distribution. In other words,
they confirmed the period-eccentricity relation and implied that close
and  wide  binaries  were  formed by  different  processes.   However,
\citet{R10} have not found any dependence of eccentricity on period at
$P>10^3$ days in  a similarly complete but larger  sample; for 82 such
binaries  in   their  sample  the  mean  eccentricity   is  0.47.  The
eccentricity  distribution appears  uniform in  the range  [0-0.6] and
falls at larger values (see their Fig.~15).

So far, studies of the eccentricity distribution were based on orbital
elements,  both spectroscopic  and  visual. Long-term  radial-velocity
monitoring can achieve  a complete census of orbits  in a given sample
\citep[e.g.][]{Griffin2012}.  Visual  orbits are often  constrained by
the limited time coverage that extends only to $\sim 200$ years in the
best cases.  While the discovery of visual binaries does not depend on
the   eccentricity,   the    orbit   calculation   does,   introducing
``computational selection'' in  the orbit catalogs \citep{Finsen1936}.
Both \citet{DM91} and \citet{R10}  used visual orbit catalogs in their
analysis  of   nearby  samples.   Based   also  on  the   catalogs  of
spectroscopic  and  visual  orbits,  \citet{Abt2006}  found  that  the
average  eccentricity of  late-type  stars increases  with period  and
saturates  at $\langle  e  \rangle =  0.52  \pm 0.02$  at  $P >  10^5$
days. These findings, however,  are likely biased by the computational
selection inherent to the orbit catalogs.

In this  paper we study  the eccentricity distribution of  wide visual
binaries by a  new method that does not rely  on orbit calculation. It
can  access  binaries with  periods  longer  than  1000 years  without
computational selection  bias. The  idea is to  use the  direction and
speed of  the observed orbital  motion. Only a  short arc needs  to be
covered  by the  data.  The  statistics of  the  observed quasi-linear
motion  depend on the  eccentricity distribution,  which then  can be
reconstructed from the observational data without computing the orbits
\citep{Tok98}.   This    method   has   been    used   previously   by
\citet{Shatsky2001}; it will benefit from precise {\it Gaia} astrometry in the near future.

The  method   is  explained  in   Section~\ref{sec:method}.   Then  in
Section~\ref{sec:data}  we present the  input data:  a sample  of 
solar-type visual binaries  within  67\,pc.   Simulations are  described  in
Section~\ref{sec:simulations},  the derived  eccentricity distribution
is given in Section~\ref{sec:res}, with the details of the restoration
technique  provided  in  the  Appendix.   The paper  closes  with  the
discussion of the results and summary in Section~\ref{sec:disc}.

\section{The method}
\label{sec:method}

\begin{figure}
\includegraphics[width=\columnwidth]{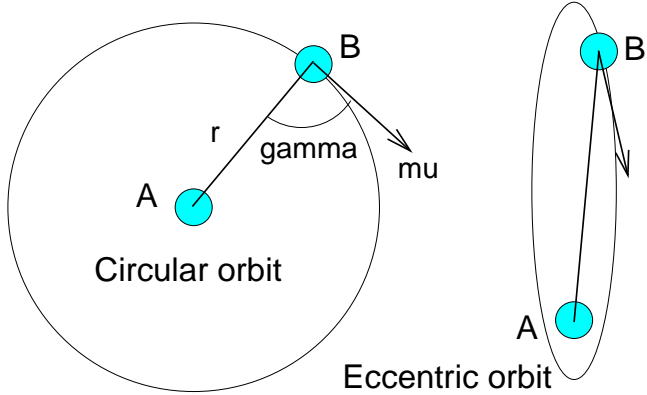}
\caption{Scheme of  the observed motion in a  binary system.  Left:
  circular orbit, right: eccentric orbit.
\label{fig:diag} 
}
\end{figure}

Figure~\ref{fig:diag} explains the idea.  When only a short segment of
the  orbit is  observed,  we can  measure  the speed  of the  relative
orbital  motion $\mu$  and the  angle $\gamma$  of the  orbital motion
relative  to the  vector joining  the binary  components.   A circular
orbit seen ``face-on'' will have $\gamma = \pm 90^\circ$, while for an
eccentric  orbit we  expect  a near-radial  motion  with $\gamma  \sim
0^\circ$ or $\gamma \sim 180^\circ$.  For binaries seen in projection
on  the sky  at random  orbital phases,  statistical  distributions of
$\gamma$ and  $\mu$ still depend on the  eccentricity. Throughout this
paper   we   assume  that   binary   orbits   are  oriented   randomly
(isotropically) relative to the line of sight. 

Here  we  extend  the  method  of \citet{Tok98}  by  considering  both
parameters,  $\gamma$ and  $\mu$.   Hereafter they  are called  linear
motion parameters  (LMPs).  The joint distribution  of $(\gamma, \mu)$
brings more  information than just  $\gamma$ alone. Note  however that
while  $\gamma$ is  a  purely geometric  parameter, the interpretation  of
$\mu$  requires  knowledge  of  parallax and  mass.   Radial  velocity
difference between the components  is yet another observable parameter
that depends on the eccentricity; it is not exploited here. 

There are several  obstacles to our project.  First,  the LMPs of wide
binaries can be  distorted by motions in subsystems.   For this reason
we base our work on the FG-67 sample of solar-type stars within 67\,pc
\citep{FG67a,FG67b}. This sample is  large enough, while the subsystem
census  is reasonably  good.  Second,  wide binaries  move  slowly and
their  motion may  not be  measurable from  the available  data.  This
limits the  largest separations and  longest periods amenable  to such
study.   Third, we must  avoid biases.   For example,  if we  use only
binaries  with  detectable motion,  a  bias  on  $\mu$ (and  hence  on
eccentricity) could  be introduced. However, removal of some  systems from the
statistical sample for lack of sufficient data is independent of their
(unknown) eccentricity, hence it should not bias the result.

\subsection{Main relations}
\label{sec:rel}

Orbits  and  periods  of wide  binaries  are  not  known. We  use  the
projected separation  $r = \rho/p$ ($\rho$ is  the angular separation,
$p$ is the parallax) as a proxy of the semi-major axis $a$.  The third
Kepler law allows a statistical estimate of the orbital period that we
denote by $P^*$, to avoid confusion with the true period $P$. So,
\begin{equation}
r^3 / (P^*)^2 = M \;\;\; {\rm or} \;\;\; P^* = r^{3/2} M^{-1/2},
\label{eq:Kepler}
\end{equation}
where $M$ is the mass sum in solar units and $P^*$ is in years.

The  characteristic  orbital velocity  is  $2  \pi  r/P^*$ AU~yr$^{-1}$.   We
multiply it by  the parallax, converting to arcsec~yr$^{-1}$,  and obtain the
characteristic speed of the relative orbital motion $\mu^*$ as
\begin{equation}
\mu^* = p (2 \pi r)/P^* = 2 \pi \rho^{-1/2} p^{3/2} M^{1/2}. 
\label{eq:mu*}
\end{equation}
Strong  dependence on  the  parallax $p$  calls  for selecting  nearby
stars. Also, close binaries move faster, but we are interested here in
wide  binaries! The  sample is  constructed by  imposing  some minimum
value of  $\mu^*_{\rm min}$. This is  equivalent to an  upper limit on
separation $ r_{\rm max}$ that depends on mass and distance,
\begin{equation}
r < r_{\rm max} = M \; ( 2 \pi p/\mu^*_{\rm min})^2 .
\label{eq:rmax}
\end{equation}

Simulations  show that  the  ratio $\mu/\mu^*$  is  a random  quantity
depending  on the  orbital phase,  projection, and  eccentricity.  Its
median value is about 0.5.  We compare the observed distributions
of  the normalised  motion $\mu'  = \mu/\mu^*$  and $\gamma$  with the
results  of  simulations.   
It can be easily proven that a bound binary system has $\mu' < \sqrt{2}$. This corresponds to the $B<1$ criterion of \citet{Pearce2015}. When the relative motion is faster than this, the double star is an optical pair \citep{Kiyaeva2008}. Considering measurement errors, we adopt a slightly relaxed criterion $\mu' < 1.5$ for physical pairs. 

\section{Observational data}
\label{sec:data}

\subsection{The sample}
\label{sec:sam}

Our main  sample consists of  solar-type binary systems  within 67\,pc
from \citep[][hereafter FG-67]{FG67a}  with projected separations $r >
50$\,AU.   The initial  selection of  886 physical  pairs  was further
filtered  by the  criterion $\mu^*  > 10$\,mas~yr$^{-1}$,  leaving 344
objects.   This  restriction  is  dictated  by  the  accuracy  of  LMP
determination from the available  data.  The masses of both components
(including known subsystems) needed for the calculation of $\mu^*$ are
taken   from  the   FG-67   compilation,  the   parallaxes  are   from
\citep{HIP2}.

\begin{figure}
\includegraphics[width=\columnwidth]{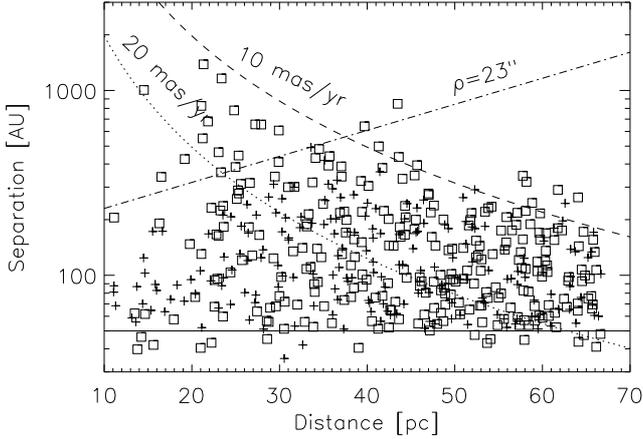}
\caption{Projected separations of  wide binaries studied here (squares
  -- main sample,  crosses -- extended sample).  The  dashed and dotted
  lines depict  upper limits  on projected separation  for $\mu^*_{\rm
    min} =10$\,mas~yr$^{-1}$ and $\mu^*_{\rm min} =20$\,mas~yr$^{-1}$, respectively,
  adopting $M = 2 {\cal  M}_\odot$.  Angular separation of $23''$ is indicated
  by the dash-dot line.
\label{fig:sample} 
}
\end{figure}

The main sample was complemented  by additional binaries from the {\em
  Hipparcos} catalog  with parallax larger than  15\,mas and projected
separation larger  than 50\,AU (minimum  separation of 0\farcs75  at a
distance of 67\,pc).  The  {\em Hipparcos} catalog lists only binaries
with  separations  up to  $23''$,  restricting  the maximum  projected
separations.   Stars  that already  belong  to  the  main sample  were
excluded from  the extension.  For  stars with masses less  than $\sim
0.8\, {\cal M}_\odot$,  {\em Hipparcos} is not complete to 67\,pc, so
the extended sample is also not complete, while still being volume-limited.

Visual magnitudes of the components  of the extended sample were taken
from the  Washington Double Star  Catalog \citep[WDS,][]{WDS}.  Masses
were estimated from absolute magnitudes in the $V$ band using standard
relations  for  main  sequence,  as in  \citep{FG67a}.   Objects  with
primary mass larger than  2\,${\cal M}_\odot$ were removed (22 total),
as their  masses are  not determined well  by the  standard relations.
Then the  characteristic speed $\mu^*$ was computed  and only binaries
with $\mu^*  > 10$\,mas~yr$^{-1}$ were kept.   This left 231  binaries in the
extended  sample.   The combined  sample  thus  contains 575  entries.
Figure~\ref{fig:sample}  shows projected  separations  versus distance
and the  selection limits  of our samples.  As explained  below, only
85\% of  binaries are included  in the statistical analysis,  the rest
having insufficient data for LMP determination. 

The  census  of  subsystems in  the  FG-67  paper  is as  complete  as
possible,  with well-known detection  limits ($\sim$80\%  complete for
subsystems in  the primary components and $\sim$50\%  for the secondary
subsystems).   In  the  extended  sample, however,  the  detection  of
subsystems  is less complete.   It was  cross-checked with  the latest
(2010)  version of  the Multiple  Star Catalog  \citep{MSC}, revealing
only 27 pairs with subsystems  out of 231.  The combined list contains
113 multiples (88  subsystems in the primary, 40  in the secondary, 15
in both).  Overall,  20\% of wide binaries in  the combined sample and
25\%  in  the  main  sample  contain known  inner  subsystems.   In  a
population simulated  according to the  prescription of \citep{FG67b},
48\%  of  binaries  with  periods  from  $10^5$  to  $10^6$  days  have
subsystems.  Orbital  periods of most subsystems are  shorter than the
time span of the data, so they should not distort the LMPs.  There are
a  few notable  exceptions, however, such as HIP~17129.   Some binaries  have  even wider
components, but this circumstance is neglected in the present work.

\subsection{Determination of linear motion}
\label{sec:amp}

Measurements of the relative positions of selected binary stars (time,
position angle, separation) were obtained from the WDS on our request.
Apart from  the {\em Hipparcos}  number, each binary is  identified by
its WDS  code based on J2000  coordinates and by  the discoverer code;
the latter discriminates between several catalog entries with the same
WDS code, e.g. resolved subsystems.

\begin{figure}
\includegraphics[width=\columnwidth]{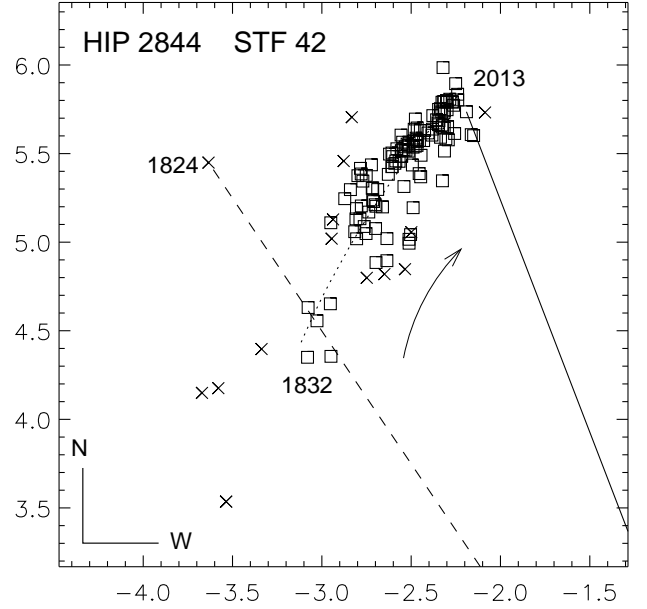}
\caption{Observed motion  of HIP~2844 (00360+2959,  STF~42~AB).  Scale
  in arcseconds,  North up, East left. Squares  mark used measurements
  (first in 1832), crosses -- rejected ones. The dashed and full lines
  connect the primary component  located at coordinate origin with the
  first and  last positions of the  companion. The dotted  line is the
  LMP trajectory.
\label{fig:HIP2844} 
}
\end{figure}

The  motion in position angle $\theta$ and separation $\rho$
was approximated by linear functions of time:
\begin{eqnarray}
\theta(t) & \approx & \theta_0 + \dot{\theta} (t - t_0) \\
\rho(t)  & \approx & \rho_0 + \dot{\rho} (t - t_0).
\end{eqnarray}

The reference time  moment $t_0$ is the average  time of observations,
as in such case the slope and mean value are not correlated.  The four
LMPs  are $(\theta_0, \dot{\theta},  \rho_0, \dot{\rho})$.   Motion of
binaries is  better represented by linear models  in polar coordinates
than  in rectangular  coordinates $X,Y$  (consider a  circular face-on
orbit for example).

The LMPs are computed from  the measured speed of relative angular and
radial motions $\mu_t = \rho_0 \dot{\theta}$ and $\mu_r = \dot{\rho}$,
so  that  $\mu  =  \sqrt{  \mu_t^2  + \mu_r^2}$  and  $\gamma  =  {\rm
  atan}(\mu_t, \mu_r)$.  The errors of  the fitted LMPs are computed in
the standard  way. We assume that   all LPMs are
statistically  independent  and  determine  the errors  of  $\mu$  and
$\gamma$ accordingly as
\begin{eqnarray}
\sigma_\mu^2 & = & \mu^{-2} \; [ (\dot{\rho}\sigma_{\dot{\rho}})^2 +  
       (\rho_0^2 \dot{\theta} \sigma_{\dot{\theta}})^2 +
       (\rho_0 \dot{\theta}^2  \sigma_{\rho_0})^2 ] , \\ 
\sigma_\gamma^2   &  = &  \mu^{-4} \; [
  (\dot{\rho}\sigma_{\dot{\rho}})^2 +
     (\rho_0 \sigma_{\dot{\rho}}\sigma_{\dot{\theta}}  )^2 +
    (\dot{\rho} \dot{\theta}  \sigma_{\rho_0})^2 ] .
\end{eqnarray}

Figure~\ref{fig:HIP2844}  illustrates  typical  data. The  dashed  and
solid lines  connect the first and last  positions, respectively, with
the primary  component located at coordinate origin  outside the plot.
The  dotted arc  is  a  linear motion  in  polar coordinates.   Points
deviating by  more than $3  \sigma$ from the  fit in either  $\rho$ or
$\theta$ were rejected iteratively  (they are plotted as crosses). All
remaining  measurements (squares)  are considered  with  equal weights.

Observations  of  some  binaries   cover a substantial  arc.   In  such
situations, we separated the data into segments with angular motion of
no more than $20^\circ$. We  also require each data segment to contain
at least three  points and to cover the time interval  of no less than
two years. The LMPs computed over the last data segment (typically the
most accurate) are taken as the final values.

The quality  and quantity of available  data is very  diverse; in some
cases  the  measurements  have  a  strong scatter  and  the  LMPs  are
determined with  large errors.   In a few  cases, one or  two strongly
deviant measurements  were deleted  manually.  For some  binaries, the
data are  not sufficient.  This happens for  recently discovered pairs
with  a  short  time  coverage  and  for  binaries  with  only  a  few
measurements of low accuracy.  Several binaries with $\rho >10''$ also
fall  in this category,  as measurement  errors usually  increase with
separation.  These pairs were  excluded from the statistical analysis.
We  also   excluded  binaries   with  errors  of   $\gamma$  exceeding
$15^\circ$.  All  binaries are physical,  as their relative  motion is
slow enough ($\mu' < 1.5$, see Section~\ref{sec:rel}).

For some  binaries in our sample  we found visual orbits  in the Sixth
Catalog   of  Orbits   of  Visual   Binary  Stars,   VB6  \citep{VB6}.
Considering  only 150 orbits  with periods  shorter that  2500\,yr, we
computed the LMPs for the same moments $t_0$ from the orbital elements
and used  those.  Comparison of ``orbital'' and  directly derived LMPs
shows  their excellent agreement,  which is  natural because  both are
based on  the same observations.  In  the extended sample,  we did not
request the observations of orbital pairs from the WDS and simply used
their orbits to compute the LMPs for $t_0 =1991.25$.

For  pairs  with  known   visual  orbits,  the  FG-67  database  lists
semi-major  axis instead of  separation.  We  removed from  the sample
orbital  binaries with  projected separations  less than  40\,AU; they
were included  by error because their semi-major  axis exceeds 50\,AU.
However, 13  orbital binaries  with separations between  40 and  50 AU
were retained.  On the other  hand, some orbital pairs with semi-major
axis less than 50\,AU  but projected separations exceeding 50\,AU were
rejected  initially, but later  recovered in  the extended  sample and
moved back to the main sample.

\subsection{Data overview and tables}
\label{sec:sumdat}

The combined  sample contains  575 wide binaries,  but only  477 pairs
with  good  data  are  left  for  statistical  analysis.   The  median
parameters of the main, extended, and combined samples are listed in
Table~\ref{tab:sample}. As expected, the extended sample has a smaller
mass of primary components and is on average located at closer
distance compared to the main sample. The last line of the Table
characterises the simulated sample (Section~\ref{sec:samp}).  

Orbital periods $P^*$ estimated  from projected separations are mostly
comprised between $10^5$ and $10^6$  days, i.e. on the right-hand side
of   the  maximum   in  the   period  distribution   at   $10^5$  days
\citep{R10,FG67b}.  As we  deal  here with  resolved visual  binaries,
their mass ratios are mostly above 0.5 (median 0.75).

\begin{table} 
\centering
\caption{Median parameters of the samples
\label{tab:sample} }
\begin{tabular}{l  ccc ccc c }
\hline
Sample & $N$ & $p$ & $M_1$ & $\rho$ & $r$ & $\mu$ & $\mu^*$ \\
       &     & mas         & ${\cal M}_\odot$ & $''$ & AU & \multicolumn{2}{c}{mas~yr$^{-1}$}  \\
\hline
Main     & 344  & 22.9  & 1.15 & 2.93 & 130.3 & 11.2 & 17.6 \\
Extended & 231  & 24.8  & 0.87 & 3.27 & 110.2 & 12.0 & 18.5 \\
Combined & 575  & 23.8  & 1.09 & 3.08 & 122.0 & 11.5 & 18.1 \\
Simulated & \ldots &20.5 & \ldots & 2.59 & 122.2 & 9.3 & 16.6 \\
\hline
\end{tabular}
\end{table}   

\begin{table*}  
\centering
\caption{Wide binaries within 67 pc (fragment)
\label{tab:main} }  
\begin{tabular}{l l l  cc cc   c l l }
\hline
HIP & 
WDS & 
Discovever & 
$p$ & 
$r$ & 
$M_1$ & 
$M_2$ & 
$\mu^*$ & 
Type & 
Tag \\
& & code & 
mas & 
AU & 
${\cal M}_\odot$ & 
${\cal M}_\odot$ &
mas yr$^{-1}$ & 
 &  \\
\hline
50  &   00006$-$5306 &HJ 5437        &     16.8 &  105.1 &   1.55 &0.87  &  16.1 &C &  M \\
96  &   00012$+$1357 &WNO 12         &     28.5 &  418.2 &   0.68 &0.64  &  10.1 &L &  E \\
110 &   00014$+$3937 &HLD 60         &     20.1 &   58.1 &   0.91 &0.84  &  22.0 &O &  E \\
169 &   00021$-$6817 &I 699 AB       &     65.2 &   64.6 &   0.61 &0.53  &  54.3 &O &  E \\
223 &   00028$+$0208 &BU 281 AB      &     23.5 &   65.0 &   1.17 &0.84  &  25.9 &L &  M \\
473 &   00057$+$4549 &STT 547 AB     &     88.4 &   68.4 &   0.61 &0.60  &  74.1 &O &  E \\
495 &   00059$+$1805 &STF 3060 AB    &     26.9 &  130.1 &   0.88 &0.79  &  19.2 &L &  M \\
522 &   00063$-$4905 &HDO 180        &     38.9 &  120.6 &   2.25 &0.54  &  37.2 &LM&  M \\
1292&   00162$-$7951 &CVN 14         &     57.2 &   67.6 &   0.95 &0.45  &  51.7 &L &  m \\
\hline 
\end{tabular} 
\end{table*}

\begin{table*}  
\centering
\caption{Linear motion parameters of binaries (fragment)
\label{tab:amp} }  
\begin{tabular}{l l r  rr rr   rrr }
\hline
HIP & 
Type & 
$t_0$, $\Delta t$ &
$\theta_0$ &
$\dot{\theta}$ &
$\rho_0$ &
$\dot{\rho}$ &
$\mu$ &
$\gamma$ &
$\mu'$, rms \\ 
& &  
yr & 
deg & 
deg yr$^{-1}$ & 
\arcsec & 
\arcsec yr$^{-1}$  & 
mas yr$^{-1}$  & 
deg &
 $^\circ /''$ \\
\hline
50   &  C & 1981.6 &321.282 &   0.372 &  1.769 &$-$0.0078 & 13.9 &  124.2 &  0.87     \\
     &    &    52.9 &  0.647 &   0.034 &  0.029 &   0.0016 &  1.2 &    5.3 &  2.5/0.11 \\
96   &  L & 1980.8 &203.971 &   0.003 & 11.897 &$-$0.0052 &  5.2 &  174.0 &  0.52     \\
     &    &   108.1 &  0.080 &   0.003 &  0.083 &   0.0029 &  3.0 &    7.2 &  0.4/0.38 \\
110  &  O & 1991.2 &176.367 &$-$0.430 &  1.171 &   0.0076 & 11.6 &$-$49.2 &  0.53     \\
169  &  O & 1991.2 &125.191 &   0.213 &  4.212 &   0.0068 & 17.1 &   66.6 &  0.31     \\
223  &  L & 1984.0 &170.567 &$-$0.349 &  1.526 &   0.0014 &  9.4 &$-$81.7 &  0.36     \\
     &    &    57.0 &  0.198 &   0.010 &  0.015 &   0.0008 &  0.8 &    4.7 &  1.0/0.07 \\ 
473  &  O & 1991.2 &178.737 &   0.417 &  6.050 &   0.0044 & 44.3 &   84.3 &  0.60     \\
495  &  L & 1949.3 &125.756 &   0.132 &  3.504 &$-$0.0019 &  8.3 &  103.4 &  0.43     \\
     &    &   155.3 &  0.062 &   0.002 &  0.008 &   0.0002 &  0.1 &    1.5 &  0.7/0.10 \\
\hline \end{tabular} 
\end{table*}

The  median   errors  of  $\mu$  and  $\gamma$   are  0.5\,mas~yr$^{-1}$  and
$2.6^\circ$,  respectively. The  actual distribution  of  those errors
closely  matches negative-exponential,  $f(\sigma_{\mu})  \propto \exp
(-\sigma_{\mu}/0.8)$    and     $f(\sigma_{\gamma})    \propto    \exp
(-\sigma_{\gamma}/3.8)$.   Note  that  the 5\%  error  on  $\mu$  and  the
$3^\circ$ error on $\gamma$ match each other.

The Referee pointed out that rejection of binaries with $\sigma_\gamma > 15^\circ$ might bias the statistics, as binaries on eccentric orbits near apastron move slowly and might be preferentially excluded from the sample by this criterion. The above exponential formula predicts that only 2 per cent of binaries would have $\sigma_\gamma$ exceeding $15^\circ$, while in reality there are 57 such pairs (five times more), suggesting that most rejections are simply worse-than-average data. A small fraction of the rejected binaries can indeed be eccentric pairs near apastron, but the bias caused by their rejection should be correspondingly small. Unfortunately, the   diversity of the data quality does not allow an easy evaluation of such bias by simulation. The median separation of the 57  pairs 
rejected by the $\sigma_\gamma > 15^\circ$ criterion is 4\farcs7, wider than in the complete sample, while their median relative motion is slower, 7~mas~yr$^{-1}$ (median $\mu'=0.36$).    

The median characteristics of 150 binaries with known orbits belonging
to the combined sample are: period 528\,yr, semi-major axis 2\farcs22,
mean eccentricity 0.53.

Table~2, available in full  electronically, lists both samples ordered
by the  Hipparcos number in Column  (1).  Column (2)  contains the WDS
code, Column (3) the  discoverer code. The following columns list
(4): parallax  in mas,  (5): projected separation  in AU, (6)  and (7):
masses of  the primary and  secondary components in solar  units, (8):
characteristic motion $\mu^*$ in mas~yr$^{-1}$. Then in Column (9) a 1-letter
code  describing the  type  of motion  is  given: L  for  a short  arc
(linear), C  for a  large and  curved arc or  a substantial  (at least
factor 1.5) change  in separation (this classification  is
subjective), and O for binaries  with known orbits.  Letter M is added
to the type for binaries  containing known inner subsystems.  The last
Column (10) has the following flags: M for the main sample with known LMP,
E for the extended sample with known LMP, m and e --- binaries from the main
and  extended  samples, respectively,  excluded  from the  statistical
analysis for lack of sufficient data.

Table~3 lists  the LMPs  of binaries  used in this  work (flags  M and
E). The Columns (1) and (2) repeat the {\em Hipparcos} number and type
of data  from Table~2.   Column (3) contains  the mean epoch  $t_0$ in
Besselian years. The following Columns  (4) through (7) list the LMPs:
$\theta_0$ (degrees),  $\dot{\theta}$ (degrees~yr$^{-1}$), $\rho_0$ (arcsec),
$\dot{\rho}$  (arcsec~yr$^{-1}$). The  LMPs  are used  to  compute the  total
motion $\mu$ given  in Column (8) in mas~yr$^{-1}$ and  the angle $\gamma$ in
Column (9),  in degrees  and with proper  sign.  The last  Column (10)
gives the  normalised motion $\mu'  = \mu/\mu^*$.  When the  LMPs were
determined here, the  following line of Table~3 gives  their errors in
the same units.  Column (3) of the second line then contains the time span of the data in years 
 and Column (10) contains  the rms scatter
from the linear elements in $\theta$ and $\rho$, in degrees and arcseconds
respectively. When the  LMPs are calculated from the  orbits (type O),
no errors are given.

The following statistical analysis uses only two parameters $\mu'$ and
$\gamma$. The angles are transformed into the interval (0,90$^\circ$),
i.e. we  set $\gamma \rightarrow  |\gamma|$ for $-90^\circ <  \gamma <
90^\circ$ and $\gamma  \rightarrow 180^\circ -|\gamma|$ otherwise. The
characteristic  speed   $\mu^*$  is  computed   with  {\em  Hipparcos}
parallaxes and estimated masses. If  the masses are biased (e.g. by not
accounting  for  additional  unknown  components), this  bias  affects
$\mu^*$ and $\mu'$. Comparison with simulations shows a good agreement
in the median $\mu'$ values, so the bias, if any, should be small.

\section{Simulations}
\label{sec:simulations}

\subsection{Simple binaries}
\label{sec:bin}

Our simulation  code generates a large  number $N$ of  binaries with a
given  eccentricity  distribution.    Orbital  elements  that  do  not
influence the  shape of  the orbit (period  $P=1$ and  semi-major axis
$a=1$) take  fixed dimensionless values.  The  longitude of periastron
$\omega$   is  uniformly  distributed   in  the   interval  $[0^\circ,
  360^\circ]$, the  inclinations $i$ are in  the $[0^\circ, 90^\circ]$
interval with a uniform distribution of $\cos i$.  The binary position
is calculated for some random moment of time and for the moment $0.005
P$ later.  The tangential and radial components of the relative motion
are computed  from the displacement  over $0.005 P$, giving  the total
speed  $\mu$  (in the  same  dimensionless  units)  and its  direction
$\gamma$ relative  to the radius-vector.  As we simulate  only systems
with direct motion ($i  \le 90^\circ$), all $\gamma$ are non-negative.
They  are ``folded''  into  the $[0^\circ,  90^\circ]$ interval.   The
normalised motion $\mu'  = \mu \rho^{1/2} /(2 \pi)  $, where $\rho$ is
the ``observed'' separation of each binary.

\begin{figure}
\includegraphics[width=\columnwidth]{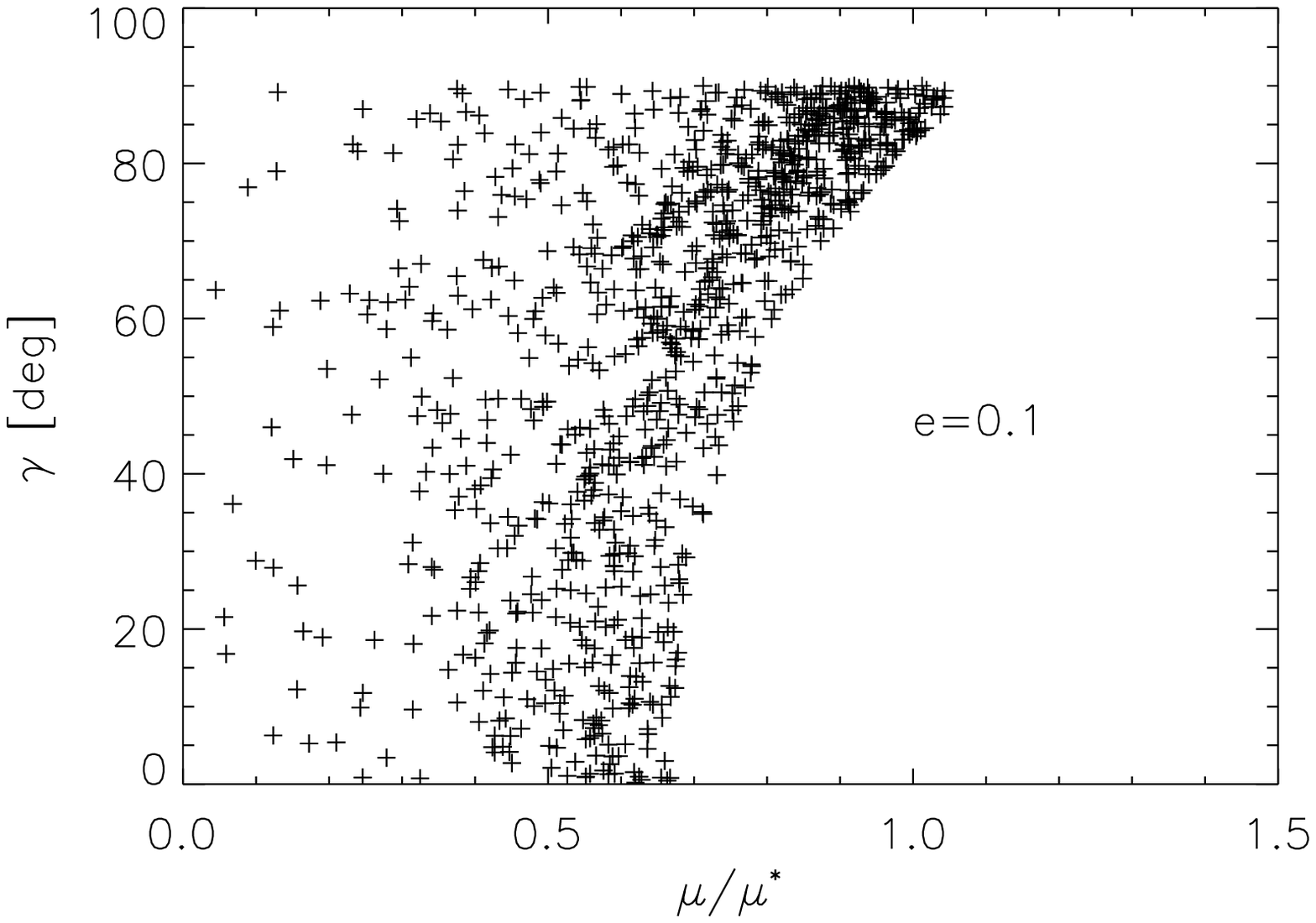} 
\includegraphics[width=\columnwidth]{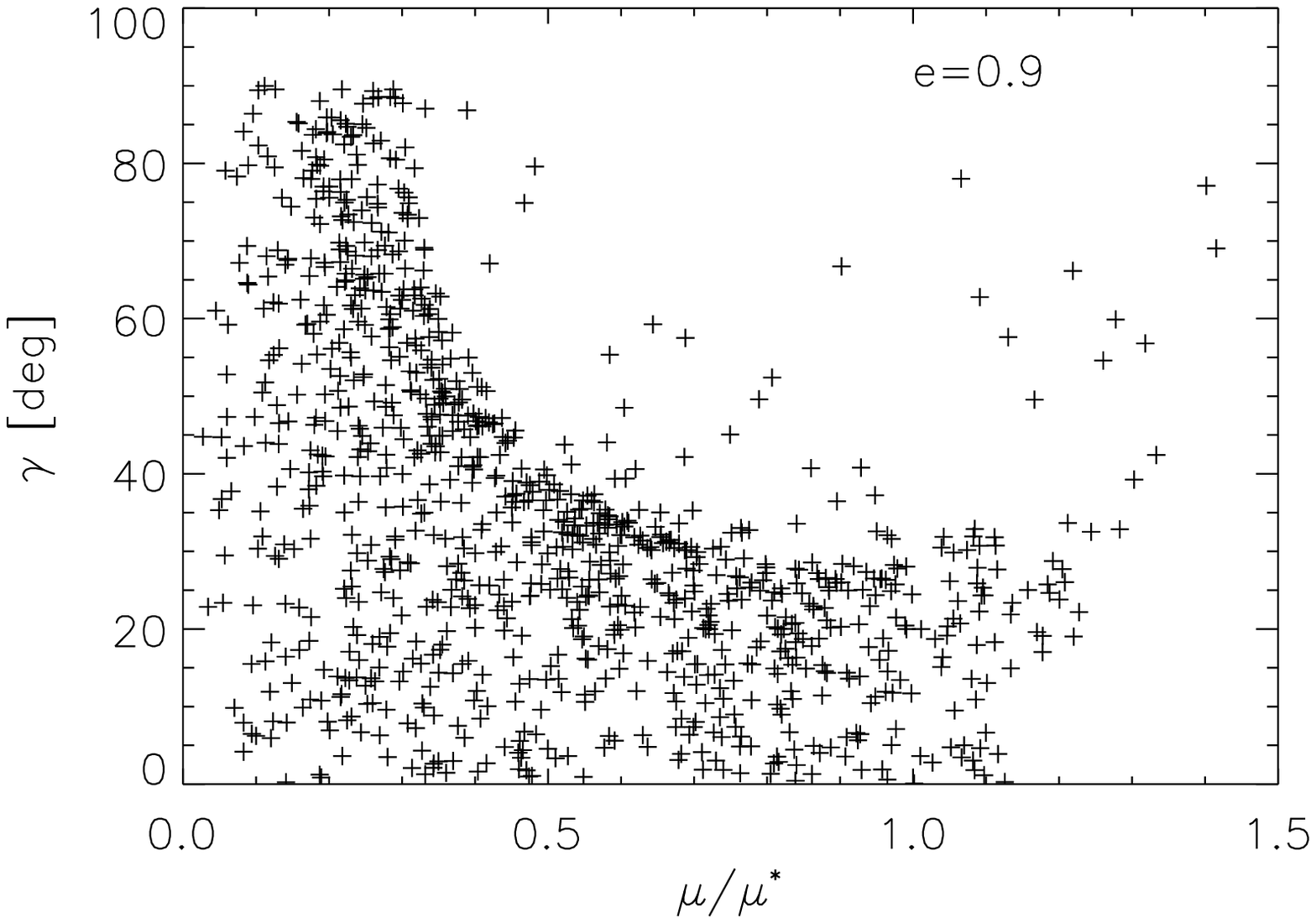} 
\caption{\label{fig:simul}  Relation between  the  normalised relative
  motion $\mu' = \mu/\mu^*$  and angle $\gamma$ for simulated binaries
  with $e=0.1$ (top) and $e=0.9$ (bottom). Each plot contains $N=1000$
  points.  }
\end{figure}

\begin{figure}
\includegraphics[width=\columnwidth]{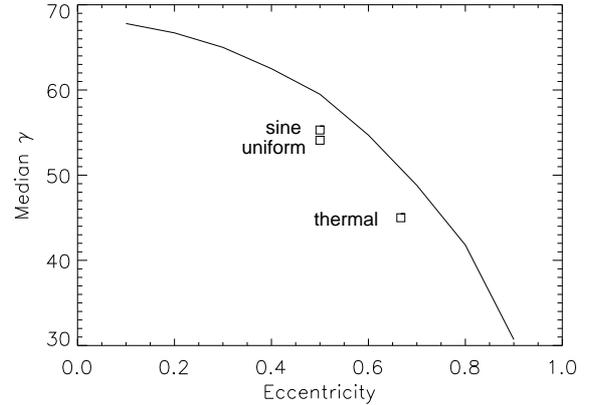}
\caption{Results of simulations for simple binaries: dependence of the
  median $\gamma$ on  the eccentricity.  The curve is  for binaries of
  fixed  eccentricity,  while squares  depict  simulated samples  with
  thermal, uniform,  and sine eccentricity  distributions, where their
  mean eccentricity is plotted.
\label{fig:e-gamma} 
}
\end{figure}

Figure~\ref{fig:simul} illustrates the  distribution of the parameters
$(\mu', \gamma) $ for a  relatively small number $N=1000$ of simulated
binaries.   Near-circular  orbits   have  a  strong  concentration  to
$(\mu',\gamma)  \sim  (1, 90^\circ)$,  despite  projection.  Owing  to
random projection of the orbits, some circular binaries have $\gamma <
90^\circ$ and move at a  slower normalized speed (remember that $\mu'$
contains the factor $\rho^{1/2}$,  meaning that orbital motion appears
slower than  expected when  the binary is  seen at  close separation).
There  is a  positive  correlation between  $\mu'$  and $\gamma$.   In
contrast, $\gamma$ decreases and  the correlation becomes negative for
eccentric  orbits.  Figure~\ref{fig:e-gamma}  shows the  dependence of
the median angle $\gamma_m$ on eccentricity.

We   demonstrate   by  simulation   that   the  thermal   eccentricity
distribution   $f(e)=2e$   corresponds   to   the   strictly   uniform
distribution of $\gamma$ which is uncorrelated with $\mu'$. Also, for
circular  orbits the  strict  inequality $\mu'  \le 1$  holds, while
$\mu' <  \sqrt{2}$  for all bound binaries.   In the real binaries, $\mu' > 1.5$ occurs due
to the LMP errors, when the binaries are optical (chance projections),
or when they  contain subsystems. We compute statistics  only for $0 <
\mu' < 1.5$, excluding a few cases with larger $\mu'$.

\subsection{Sampling biases}
\label{sec:samp}

We studied  by simulation the  influence of our sampling  criteria and
measurement errors  on the $(\mu',  \gamma)$ statistics. To do  so, we
simulated a realistic population  of binaries distributed uniformly in
a volume of  67-pc radius.  The periods are  drawn from the log-normal
distribution  of  \citet{R10} restricted  to  the  range  from 100  to
$10^5$\,yr. The  mass sum of 2\,${\cal M}_\odot$  is assumed. Binaries
are  simulated  as  before,  the  parameters $\mu$  and  $\gamma$  are
distorted  by observational  errors  that match  the  real ones.   The
statistics  are   calculated  for  the  whole   population  (which  is
equivalent to the results for  simple binaries) and for the sub-sample
with  projected  separation above  50\,AU  and  $\mu^* >  10$\,mas~yr$^{-1}$.
These criteria reject very  close and very wide binaries, respectively
(Figure~\ref{fig:sample}).   About  45\%  of  the  simulated  population
passes the selection criteria.  In the following we use the results of
these more realistic sample simulations.

After selection, the simulated sample is a close match to the real one
(see the last line  in Table~\ref{tab:sample}). The median parallax is
a  little less  than in  the  main sample,  reflecting its  $\sim$10\%
incompleteness  \citep{FG67a}.   Median  parameters of  the  simulated
sample do not depend on the adopted eccentricity distribution.

We found that the sample selection  does not bias the median values of
$\mu'$ and $\gamma$ in any significant way. However, their correlation
can be reduced  in absolute value  by as much as a factor  of $\sim$2. 

Table~\ref{tab:sim}   quantifies    the   simulation   results   using
$N=10\,000$  simulated binaries  for  each case.   It  lists the  {\em
  median} values  of the main  parameters $\mu'_m$ and  $\gamma_m$ and
their correlation coefficient $C$.  As the eccentricity increases, the
normalised  motion  becomes  slower,  $\gamma_m$  decreases,  and  the
correlation becomes negative. The right-hand part of the Table reports
results  of  sample simulations  that  account  for  biases. The  sine
eccentricity distribution is $f(e) = \pi/2 \sin(\pi e)$.

\begin{table}
\centering
\caption{Results of simulations and observed parameters
\label{tab:sim} }

\begin{tabular}{l c c c  ccc}
\hline 
Eccentricity & 
$\mu'_m$ & 
$\gamma_m$ & 
 $C$ & 
 $\mu'_m$ & 
 $\gamma_m$ & 
 $C$  \\ 
         & \multicolumn{3}{c}{Simple binaries} &
\multicolumn{3}{c}{Sample simulation} \\
\hline
$e=0$   & 0.677 & 68.3  & 0.61    & 0.678 & 67.4 & 0.26  \\
$e=0.3$ & 0.660 & 65.7  & 0.43    & 0.674 & 65.9 & 0.19  \\
$e=0.6$ & 0.577 & 53.1  & 0.03    & 0.590 & 55.0 & 0.02  \\
$e=0.9$ & 0.437 & 31.2  & $-$0.44 & 0.432 & 30.5 & $-$0.41  \\
Uniform  & 0.606 & 54.1  & 0.18   & 0.608 & 53.5 & 0.08 \\
Sine    & 0.609 & 55.3  & 0.15   & 0.610 & 56.0 & 0.07 \\ 
Thermal & 0.546 & 45.0  & 0.00    & 0.549 & 44.7 & 0.00 \\ 
\hline
Combined (477) &  \ldots  & \ldots  & \ldots &  0.58  & 48.6  & 0.11 \\
Main (282) & \ldots  & \ldots  & \ldots      &  0.57  & 47.4  & 0.07 \\
Multiples (92) &  \ldots  & \ldots  & \ldots & 0.65   & 56.6  & 0.25 \\
Binaries (385) &  \ldots  & \ldots  & \ldots & 0.57   & 47.5  & 0.08  \\
\hline \end{tabular} 
\end{table}

\section{Results}
\label{sec:res}

\begin{figure}
\includegraphics[width=\columnwidth]{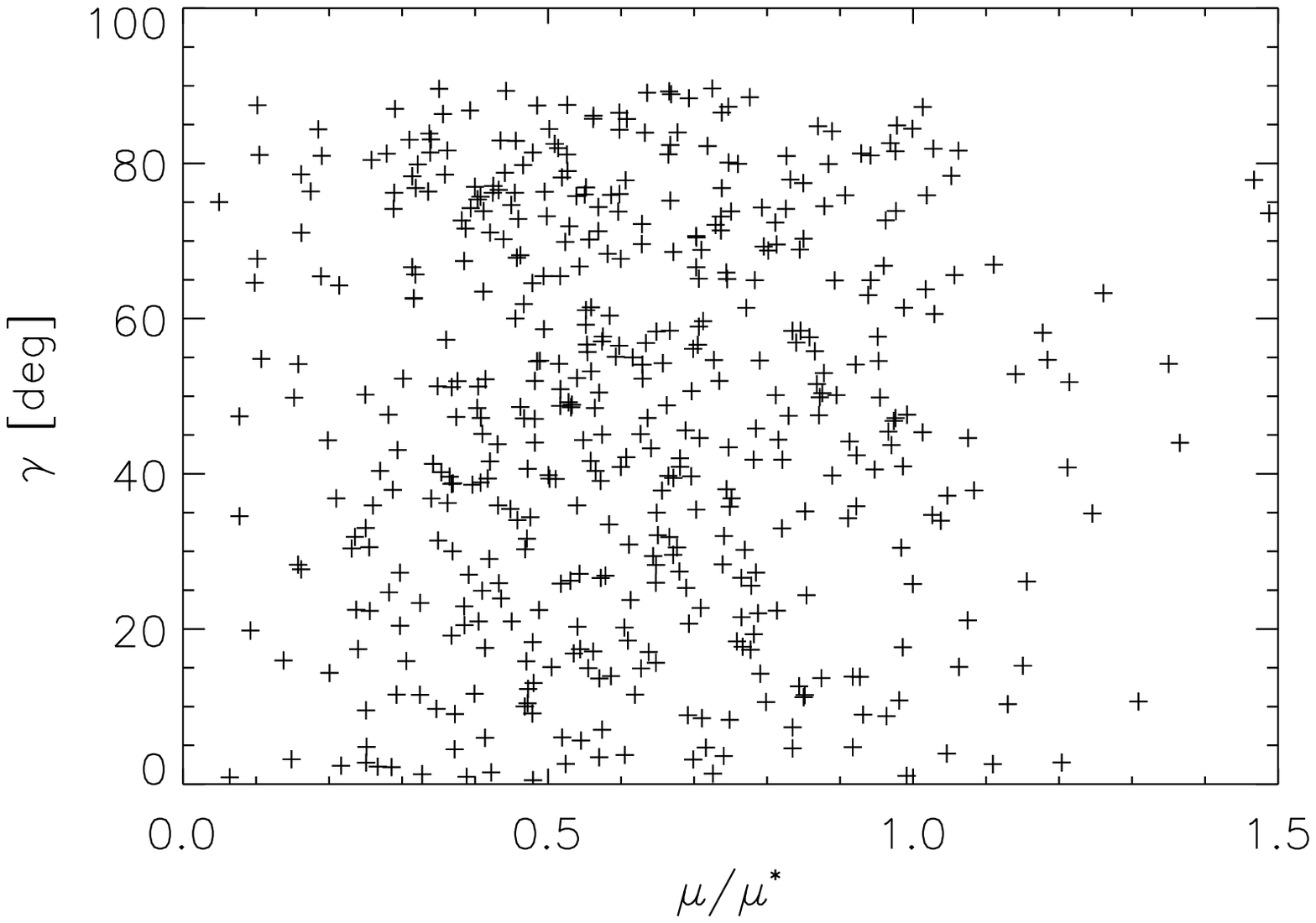} 
\includegraphics[width=\columnwidth]{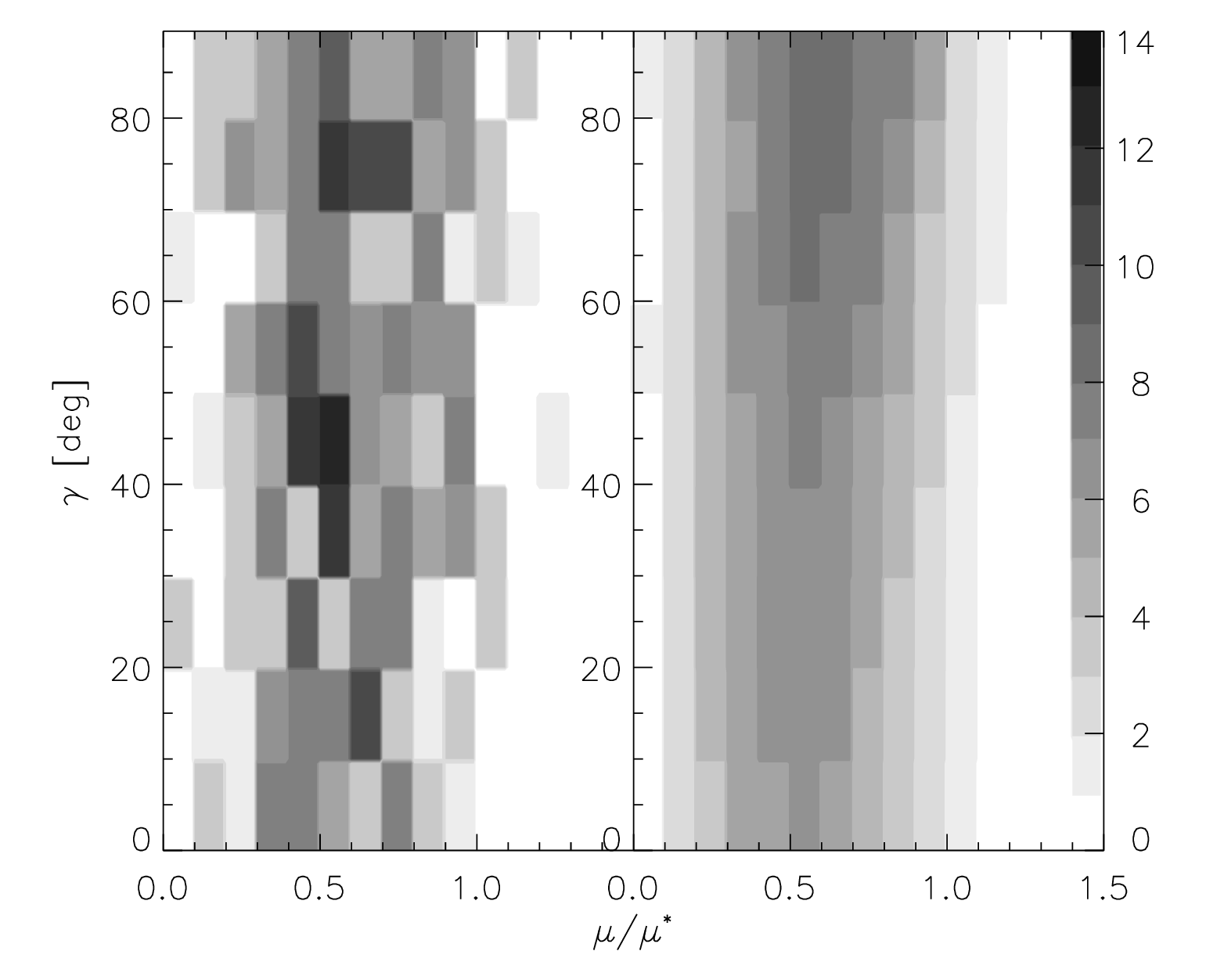} 
\caption{Top: relation between $\mu'$ and $\gamma$ for 475 binaries in
  the combined sample. Bottom: the  grayscale histogram of the data in
  15$\times$9 bins on the left is  compared to the fitted model on the
  right.  The  scale bar on  the right shows  data counts in
  the histogram.
\label{fig:gamma-mu} 
}
\end{figure}

Figure~\ref{fig:gamma-mu} (top) shows  the observed values  of $\mu'$
and $\gamma$.  The median parameters  and correlation are given in the
last  lines   of  Table~\ref{tab:sim}  for   direct  comparison  with
simulations. 

The 5-bin eccentricity distribution  was determined by the regularised
least-squares method described in the Appendix.   Briefly, the two-dimensional histogram of $\mu'$, $\gamma$ is approximated by a linear combination of five histograms produced by simulations, where each simulated histogram corresponds to the eccentricity comprised in one bin. 
The result is a set of five numbers $f_k$ representing fractional content of each bin, with 
 $\sum_k f_k = 1$. The lower part of Figure~\ref{fig:gamma-mu} illustrates how the real two-dimensional histogram is matched by its model. 

The errors of $f_k$ are estimated by the
bootstrap  technique.   We  generated   50  artificial  data  sets  by
selecting randomly the same number  of binaries from the original data
(i.e.   some  data  are  skipped,  some appear  several  times).   The
distribution was determined  for each set and averaged,  while its rms
scatter estimates the errors.   The average eccentricity is determined
more accurately than the distribution itself. Errors delivered by bootstrap were confirmed by analysis of simulated samples, thus verifying the method. 

\begin{figure}
\includegraphics[width=\columnwidth]{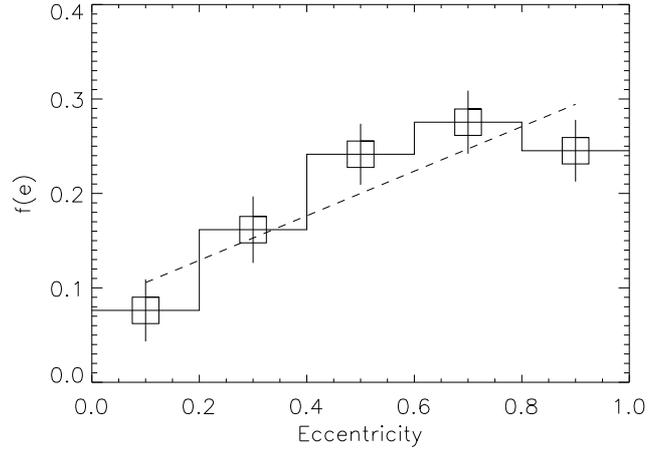}
\caption{Eccentricity  distribution of  wide binaries  and  its errors
  determined by bootstrap for the combined sample.  The dashed line is
  a linear  model $f(e) = 1.2 e + 0.4$.  
\label{fig:eplot} 
}
\end{figure}

Table~\ref{tab:res} lists the  resulting 5-bin distributions and their
errors for the main and  combined samples and various sub-samples. Two
binaries  with $\mu'>1.5$  are not  considered, reducing  the combined
sample size from 477 to 475.  The derived eccentricity distribution is
plotted in  Figure~\ref{fig:eplot}.  A linear  model $f(e) =  2 f_{\rm
  lin} e + 1 - f_{\rm lin}$ (sum of thermal and uniform distributions)
was also fitted to the $(\mu', \gamma)$ data; we give $f_{\rm lin}$ in
the right column of Table~\ref{tab:res}.

To study the dependence of  eccentricity on separation, we divided the
combined  sample  in  two  approximately equal  parts  with  projected
separations less  or larger than 100\,AU.   The average eccentricities
are $0.56 \pm 0.03$ and $0.62\pm0.03$, respectively, and the
eccentricity distribution becomes closer to a thermal one with
increasing separation.  The median mass ratios in these two groups are comparable, 0.77 and 0.72.

There  is evidence  of  different eccentricity  distributions in  pure
binaries  and  binaries   containing  subsystems.   This  matches  the
difference in the median angles  $\gamma$ between these  sub-samples (see
Table~\ref{tab:sim}).  The eccentricity distribution reconstructed for
92 multiples  in the combined  sample has a  strong peak in  the third
bin $e=$[0.4-0.6], $f_3 = 0.40$.   The average eccentricity of
multiples is  $0.52\pm 0.05$, significantly less than  for binaries in
general.    Remember  that   many   ``binaries''  contain   undetected
subsystems, so  the actual  difference may be  even larger  than found
here.

\begin{table}
\centering
\caption{Derived eccentricity distributions}
\label{tab:res}
\begin{tabular}{l  c cccc c c }
\hline
Sample  & $f_1$  & $f_2$  & $f_3$  & $f_4$  & $f_5$  & $\langle e  \rangle$ & $f_{\rm lin}$ \\
$(N)$ & \\
\hline
Combined & 0.08       & 0.16 & 0.24 & 0.28 & 0.25 &    0.59 &  0.59 \\ 
~~(475)  & $\pm$0.03  & 0.04 & 0.03 & 0.03 & 0.03 &    0.02  & \\
Main      &  0.04     & 0.19 & 0.24 & 0.23 & 0.29 &    0.62 &  0.73 \\ 
~~(280)   & $\pm$0.03 & 0.04 & 0.05 & 0.05 & 0.05 &    0.02  &      \\
Extended  & 0.14      & 0.14 & 0.20 & 0.33 & 0.19 &    0.56  & 0.39 \\
~~(195)   & $\pm$0.06 & 0.06 & 0.05 & 0.06 & 0.06 &    0.03  & \\
Multiples & 0.12      & 0.17 & 0.40 & 0.15 & 0.17 &    0.52 &  0.04 \\ 
~~~(92)   & $\pm$0.08 & 0.08 & 0.10 & 0.08 & 0.08 &    0.05  & \\
$r<100$AU & 0.13      & 0.15 & 0.21 & 0.31 & 0.19 &    0.56 &  0.37 \\ 
~~~(218)  & $\pm$0.05 & 0.05 & 0.06 & 0.07 & 0.05 &    0.03  & \\
$r>100$AU & 0.03      & 0.18 & 0.26 & 0.23 & 0.30 &    0.62 &  0.75 \\ 
~~~(256)  & $\pm$0.04 & 0.05 & 0.05 & 0.05 & 0.04 &    0.03  & \\
\hline
\end{tabular}
\end{table}

\section{Discussion}
\label{sec:disc}

The eccentricity distribution of  wide binaries is ``softer'' than the
thermal one, containing more orbits  with $e<0.2$ and less orbits with
$e>  0.8$.   To  our   knowledge,  this  is  the  first  observational
determination of $f(e)$  for wide binaries.  The {\it  ad hoc} initial
eccentricity    distribution   adopted    in    $N$-body   simulations
\citep[e.g.][their  Figure 2]{Marks2011} is  closer to  $f(e)=2e$ than
the real one.  These  authors found that the eccentricity distribution
is not affected by dynamical evolution of binaries in a cluster.  Some
binaries  are  simply  disrupted,  while the  remaining  binaries  are
unchanged. This  means that $f(e)$  for field binaries  reflects their
formation mechanism.

Appreciable difference in the eccentricities of wide binaries with and
without  subsystems   is  a  new  result,  confirming   the  work  of
\citet{Shatsky2001}.  In the hindsight,  such a difference is expected
because  very eccentric  outer  orbits are  not  allowed by  dynamical
stability  of multiple  systems.  Using  the  statistical multiplicity
model proposed  by \citet{FG67b}, we generated  a synthetic population
of  triples with  outer  periods  from 250  to  8000 years  containing
subsystems with period ratios $P_{\rm OUT}/P_{\rm IN} > 4.7$ in one of
their  components.   Assuming  that  the  outer  orbits  have  thermal
eccentricity distribution,  we removed triples that do  not conform to
the dynamical stability criterion by \citet{MA02},
\begin{equation}
P_{\rm OUT}(1-e)^{1.8}(1+e)^{-0.6}  > 4.7 P_{\rm IN} ,
\label{eq:stab}
\end{equation}
and found  that the remaining outer systems  have average eccentricity
of  $\sim$0.55,  in qualitative  agreement  with  the actual  multiple
systems.  We provide  this crude estimate only as  an illustration because
its underlying assumptions are unrealistic. It is more likely that formation mechanisms of multiple systems favour moderate eccentricities in outer orbits.

\begin{figure}
\includegraphics[width=\columnwidth]{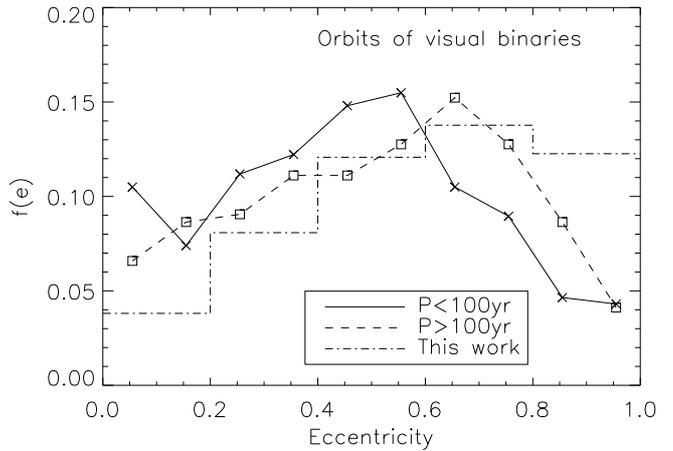}
\caption{Distribution of  eccentricity  in the VB6 for orbits
  of grade  3 or  better. Full  line: 581 orbits  with $P  < 100$\,yr,
  dashed line: 243 orbits with $P>100$\,yr. Average eccentricities are
  0.45 and  0.51, respectively.  The dash-dot line  is our  result for
  the combined sample normalised for bin size of 0.1.
\label{fig:vb6} 
}
\end{figure}

\begin{figure}
\includegraphics[width=\columnwidth]{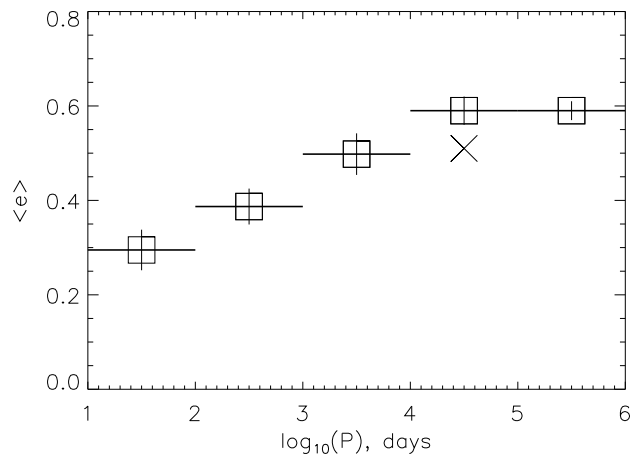}
\caption{Period-eccentricity relation.  Horizontal lines depict period
  ranges  and  average  eccentricity  in  these  ranges  according  to
  \citet{Udry1998}, \citet{Griffin2012}, \citet{Abt2006} (cross, underestimated),
  and this work (the last two period intervals). The errors are shown by vertical lines. 
 \label{fig:pe}}
\end{figure}

It  is instructive  to compare  our results  with the  distribution of
eccentricities in known  visual orbits. We selected orbits  of grade 3
or better  from the  VB6 catalog \citep{VB6}  and found  a bell-shaped
distribution   of   $e$   declining   towards   large   eccentricities
(Figure~\ref{fig:vb6}).  There  is a tendency  of smaller eccentricity
at shorter orbital periods. For  581 orbits with $P < 100$\,yr (median
period  27\,yr) we  find $\langle  e \rangle  = 0.45$,  while  for 243
orbits with  $P>100$\,yr (median period 158\,yr) $\langle  e \rangle =
0.51$. Orbits determined from a short observed arc often have a strong
positive  correlation between period  and eccentricity,  so it  is not
clear to  what extent  the VB6 trend  of eccentricity  increasing with
period is genuine. 

Comparing the  eccentricity distribution of $P>100$\,yr  orbits in VB6
with the distribution derived here,  we note their similarity up to $e
\sim  0.8$ and  the strong  deficit of  larger eccentricities  in VB6.
This  is  most  certainly   related  to  the  computational  selection
discussed by \citet{Finsen1936}. There are many binaries in our sample
with nearly  radial motion and substantial time  coverage, but without
computed orbits.  The mean  eccentricity of visual binaries derived by
\citet{Abt2006},  \citet{R10} and  others  should be  biased to  lower
values by this effect.

We note that VB6 contains a heterogeneous sample of binaries dominated
by  nearby   low-mass  stars.   The  errors  and   biases  of  orbital
eccentricities  in  VB6 are  unknown.   The  method  used here  avoids
uncertainties and biases associated with orbit calculation.

The  computational selection is  less of  a problem  for spectroscopic
binaries.   Very long  time  coverage  of binaries  in  the Hyades  by
\citet{Griffin2012}  is  particularly  useful  in this  respect.   His
Table~1 contains 29  orbits with $P>10^3$ days (and  mostly $P < 10^4$
days)  with  a  nearly  uniform  eccentricity  distribution  and a  mean
eccentricity  of 0.498$\pm$0.044.   On the  other hand,  the average  eccentricity of
solar-type binaries with  periods from 10 to $10^3$  days in the field
and in open  clusters is less,  0.31 \citep{DM91}.  Using  more extensive
data of Figure~3 from \citet{Udry1998}, we compute $\langle e \rangle =
0.295 \pm 0.043$ for $10 < P < 100$ days and $\langle e \rangle = 0.387 \pm 0.038$ for $100 <
P < 1000$ days for a total of 198 orbits in these period intervals.  

To probe  the eccentricity distribution at intermediate periods, we selected
210 visual binaries  with projected separations from 10  to 50 AU from
the main 67-pc sample, covering approximately the period range $10^4 <
P < 10^5$  days, and repeated the statistical  analysis done for wider
binaries.  In  this group, 155  binaries have known orbits.   To avoid
confusion, we do not provide  more details of this supplementary study
and only  use the  derived average eccentricity  $\langle e  \rangle =
0.59 \pm 0.03$ in the discussion.

The  dependence   of  average  eccentricity  on   period  is  presented
graphically  in Figure~\ref{fig:pe}.  At  orbital periods  longer than
$10^2$  days the eccentricity  can be  as high  as 0.9,  meaning that
tidal circularisation  is not important. Still,  the mean eccentricity
continues to increase with period.

Main results of this study can be summarised as follows:

\begin{enumerate}
\item
Eccentricities of  wide (median separation $\sim$120\,AU)  nearby (within 67\,pc)
low-mass binaries in the field  are distributed as $f(e) \approx 1.2 e
+ 0.4$, with  $\langle e  \rangle = 0.59 \pm 0.02$. 

\item
We  found marginal  evidence of  average  eccentricity  increasing with  binary
separation. Comparison  with spectroscopic binaries in  the field puts
this increase beyond doubt, confirming the period-eccentricity trend.

\item
Visual    binaries    with    high   eccentricities    are    strongly
under-represented  in the current  orbit catalogs,  biasing statistics
derived from the catalogs.

\item
Eccentricities   of   wide    binaries   containing   subsystems   are
significantly less than for the rest of the sample, $\langle e \rangle
=  0.52 \pm  0.05$.  This can  be  explained, at  least partially,  by
dynamical  stability of  multiple  systems that  does  not allow  very
eccentric outer orbits.

\end{enumerate}

\section*{Acknowledgements}

This work  used binary-star measurements collected in
the  Washington Double  Star  Catalog maintained  at  USNO.  We  thank
W.~Hartkopf for extracting the measurements on our request and stress
the usefulness  of keeping  and updating the WDS  database. 
Thoughtful comments by the anonymous Referee are much appreciated. 
 We  used the
SIMBAD   service   operated  by   Centre   des  Donn\'ees   Stellaires
(Strasbourg,   France)   and   bibliographic   references   from   the
Astrophysics Data System maintained by SAO/NASA.
%



\appendix

\section{Reconstruction of the eccentricity distribution}
\label{sec:rec}

\begin{figure}
\includegraphics[width=\columnwidth]{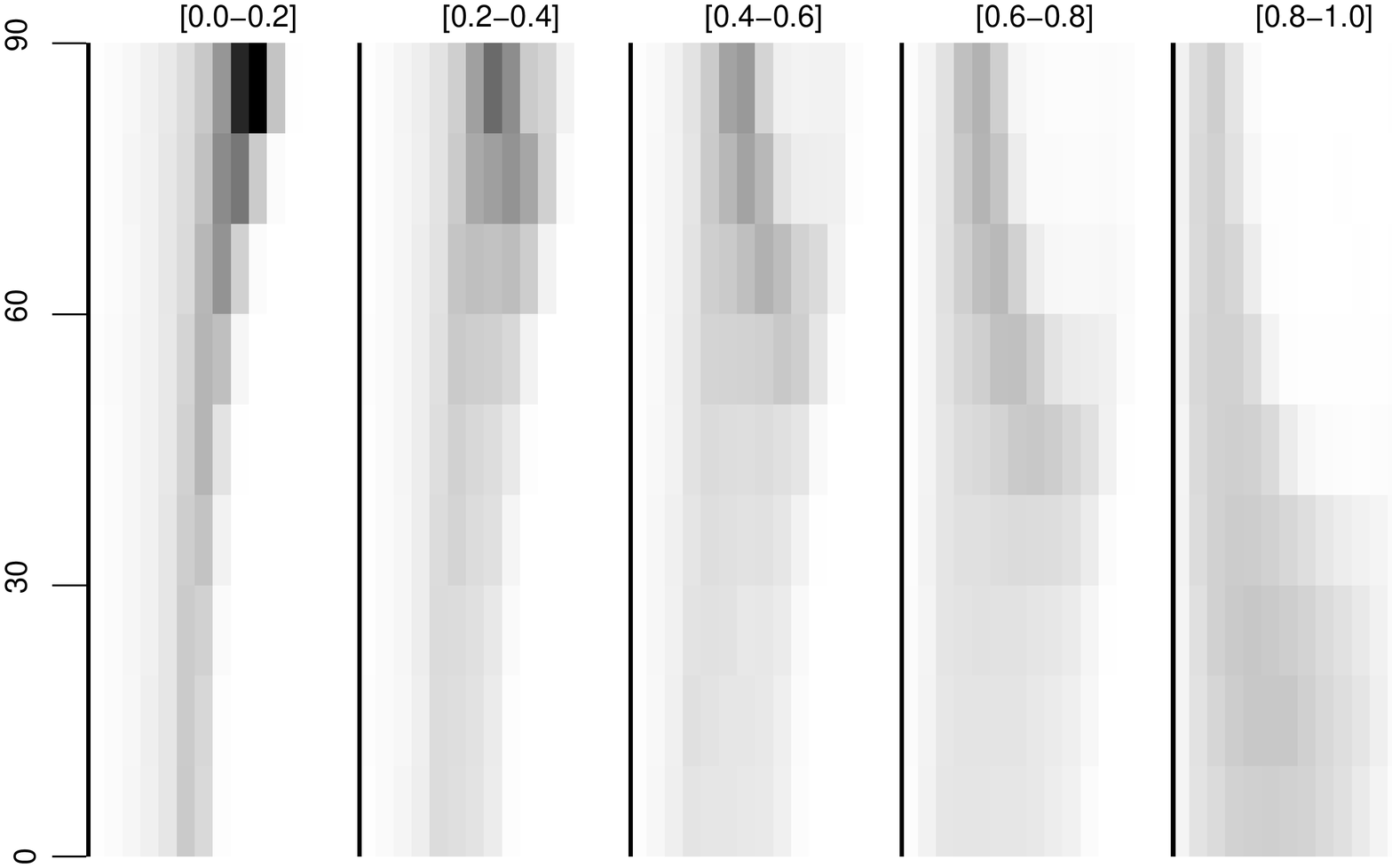} 
\caption{Grayscale representation of the $(\mu', \gamma)$ probability
  distributions.   Left  to  right:  eccentricity  intervals  [0-0.2],
  [0.2-0.4], etc. The horizontal range of each panel in $\mu'$ is from
  0 to 1.5, the vertical range in $\gamma$ is from 0 to
  90$^\circ$. The intensity scale is equal for all panels. 
\label{fig:mosaic} 
}
\end{figure}

The observed 2-dimensional distribution of $(\mu', \gamma)$ is binned on
a grid with  a step of 0.1 in $\mu'$ and  $10^\circ$ in $\gamma$, i.e.
in the 15$\times$9 array.  The  number of binaries in each grid cell is
normalised  by   $N$,  converting   the  result  in   the  probability
distribution.

The set of  five basis functions is produced by  simulating $N = 10^5$
binaries  with eccentricities distributed  uniformly in  the intervals
[0-0.2], [0.2-0.4], etc., including  the sample selection criteria and
measurement errors. The result is  binned on the same 15$\times$9 grid
and normalised to  make the sum over the grid  equal one.  These basis
functions are shown in Figure~\ref{fig:mosaic}.

Using these $k=5$ simulated  2-dimensional distributions as basis functions,
we can find  the fractions of eccentricity in  each bin $f_k$, $k=1,2,
\dots 5$  that match the  real sample.  The  $15 \times 9 =  135$ grid
cells are  numbered sequentially by the  index $i$, the  values $b_i =
n_i/N$ are fractions of the sample  in each cell. Let $g_{i,k}$ be the
basis functions  derived by simulation.   Then the problem is  to find
the 5-element vector $\mathbf{f}$  that minimises the difference
\begin{equation}
\sum_i  \left( \sum_{k=1}^5 g_{i,k} f_k  - b_i \right) ^2 
+ \alpha \sum_{k=1}^5 f_k^2    = \mathrm{min} .
\label{eq:min}
\end{equation}
A regularisation term with  a small parameter $\alpha$ is introduced
in the left-hand side.  Solution  of the inverse problem amplifies the
statistical  noise.   This  can  be improved  by  imposing  additional
condition  that the distribution  of eccentricity  should be  a smooth
function, so we added its variance with a small coefficient
$\alpha$.

Equation~\ref{eq:min}  with  additional  constraint  $\sum  f_k=1$  is
solved  by  the standard  method  of  Lagrange  multipliers. The  term
$\lambda( \sum f_k-1)$ is added  to the left hand and partial derivatives
over  $f_k$ and $\lambda$  are equated  to zero,  leading to  a linear
system with 6 unknowns $f_k, \lambda$. 

We increased  $\alpha$ until  it started to  affect the  residuals and
found that  $\alpha=0.1$ is too large  and tends to  produce an almost
uniform  distribution,  while  $\alpha   =  3\,  10^{-3}$  is  a  good
compromise. We used this value in the data processing and simulations.

Strictly  speaking, Equation~\ref{eq:min}  should  incorporate weights
inversely proportional  to the Poisson variance in  each bin. However,
for small numbers the  Poisson distribution is asymmetric, whereas the
least-squares method  assumes normally distributed  errors; this leads
to  a  potential  bias.   A statistically  rigorous  treatment  should
maximize the  likelihood function, but defining such  a function using
simulated samples is non-trivial.  The histogram matching is a simpler
practical alternative.  Yet another  reason why rigorous techniques do
not always perform as expected is the mismatch between the assumptions
(model) and the  reality.  In this study, the data  can be affected to
some extent by undetected subsystems.  The unweighted least squares is
relatively insensitive to  a small number of deviant  data points.  We
found  that  the  residual  difference  between the  real  and  fitted
histograms is compatible with the Poisson statistics, meaning that the
least-squares model adequately represents the data.

The  restoration method  was extensively  tested on  simulated samples
with  various $f(e)$.   Table~\ref{tab:test} compares  the  results of
restoration using  20 independent simulated  samples with $f(e)  = 2e$
and $N=475$ each.  The average retrieved parameters $f_k$ are close to
their true  values, while their rms  scatter is similar  to the errors
estimated  by bootstrap  (compare to  Table~\ref{tab:res}).   When the
weights proportional to $1/(1 + n_i)$ are applied to mimic the Poisson
statistics  and  to  emphasize  bins  with small  counts  $n_i$,  they
introduce  additional random noise  and the  scatter of  the resulting
$f_k$  increases (see  the  last two  lines of  Table~\ref{tab:test}).
Quasi-Poisson weights based on the  modeled counts instead of the true
$n_i$ avoid the noise amplification,  but the result might then become
biased by the $f_k$ priors used for the weight calculation.

Note that  the parameters $f_k$  are non-negative.  We could  use this
additional  constraint  and   apply  the  non-negative  LS  technique.
However,  the  result  would  then  be  positively  biased  by  random
fluctuations.  For this reason we prefer the linear LS.  We also tried
regularisation by singular value  decomposition of the matrix, keeping
the  three  largest  singular   values  in  its  inversion; the
$\alpha$-regularization gives a lower noise.

\begin{table}
\centering
\caption{Test of the  algorithm for $f(e)=2e$}
\label{tab:test}
\begin{tabular}{l  c cccc  }
\hline
Case  & $f_1$  & $f_2$  & $f_3$  & $f_4$  & $f_5$   \\
\hline
True      & 0.04     & 0.12 & 0.24 & 0.28 & 0.36 \\
No weight & 0.03     & 0.13 & 0.21 & 0.28 & 0.36 \\
          & $\pm$0.02 & 0.05 & 0.06 & 0.05 & 0.04 \\
Weighted  & 0.06      & 0.14 & 0.23 & 0.28 & 0.28 \\
          & $\pm$0.08 & 0.09 & 0.14 & 0.11 & 0.09 \\
\hline
\end{tabular}
\end{table}

Instead of  modelling the eccentricity  distribution by 5 bins,  we can
describe it by a linear function $f(e)  = 2 f_{\rm lin} e + 1 - f_{\rm
  lin}$   or  by  a   combination  of   thermal,  uniform,   and  sine
distributions   (like  a  linear   or  quadratic   polynomial).   This
alternative model requires  only two or three basis  functions and has
one or two  unknown parameters. It was also  tested by simulations and
fitted to the real data.

\bsp	
\label{lastpage}

\end{document}